\documentclass[preprint,12pt,authoryear]{elsarticle}

\usepackage{amssymb}
\usepackage{amsmath}
\usepackage{enumerate}
\usepackage{mathrsfs}

\usepackage{color} 
\usepackage{xspace}

\journal{Astroparticle Physics}

\begin{document}
\begin{frontmatter}

\title{Hydrodynamic model of nonthermal emission from the Fermi bubbles}
\author[affil:lpi]{V.~A.~Dogiel\corref{mail:dogiel}}

\author[affil:lpi]{D.~O.~Chernyshov}

\author[affil:lpi]{T.~S.~Fatekhov}

\author[affil:lpi]{A.~M.~Kiselev}

\author[affil:ncu2]{C. M. Ko\corref{mail:ko}}

\affiliation[affil:lpi]{organization={I.~E.~Tamm Theoretical Physics Division of P.~N.~Lebedev Physical Institute},
             city={Moscow},
             postcode={119991},
             country={Russia}}

\affiliation[affil:ncu2]{organization={Institute of Astronomy, Department of Physics and Center for Complex Systems, National Central University},
             city={Zhongli Dist., Taoyuan City},
             country={Taiwan (R.O.C.)}}
			
\cortext[mail:dogiel]{dogiel@td.lpi.ru}
\cortext[mail:ko]{cmko@gm.astro.ncu.edu.tw}

\begin{abstract}
We suggest a model of Fermi Bubbles (FBs) in the Galactic halo  of the altitude about $7-8$ kpc,
which is seen in non-thermal microwave and $\gamma$-ray ranges.
It was assumed that this emission is generated by relativistic electrons of cosmic rays whose origin
is still under debate. It has been assumed that the FB shell is generated in the halo by the release of energy,
generated by the routine capture of stars at the central black hole of the Galactic Centre (GC).
In this case cosmic ray electrons (CR) in the shells of the FBs of sufficiently high energies are generated by the standard shock acceleration.
However, one of the problems of this model is that the Mach number of the FB shock is not high enough to generate the observed non-thermal radiation from the halo.

We propose an alternative model of stochastic CR acceleration by Rayleigh-Taylor (RT) instabilities in the shell of the FB
at the late stages of the evolution of the shell in the halo.
Unlike the shock model of CR  acceleration, the RT model of in-situ acceleration in the FBs does not require strong shock fronts.
In our model, we derived the spectrum of RT instabilities and estimate the spectra of kinetic equations for MHD-fluctuations needed for acceleration of CRs.
We assessed the time of CR electron acceleration up to TeV energies that needed to interpret the observed data of $\gamma$-ray and microwave emission from the envelope of FBs.

\end{abstract}

\begin{keyword}
Galactic Center \sep Fermi Bubbles \sep central black hole \sep star disruptions \sep MHD turbulence
\sep Rayleigh-Taylor instability \sep cosmic ray scattering \sep non-thermal radiation mechanisms
\end{keyword}

\end{frontmatter}

\section{Introduction}


Two enigmatic $\gamma$-ray features in the Galactic central region, known as the Fermi Bubbles (FBs),
were discovered in 2010 from the Fermi-LAT data of $\gamma$-ray in the range of $1\sim 100$ GeV
and total luminosity of $F_\gamma \sim 4\times 10^{37}$ erg s$^{-1}$
\citep[see][]{su10, acker14}.

The bubble structure in the GC was also revealed in the range of microwaves which coincides nicely with that of $\gamma$-rays \citep[][]{planck13}.
The emission is in the range $23\sim 61$ GHz, and the luminosity is $\Phi_\nu\approx 1\sim 5\times 10^{36}$ erg s$^{-1}$.
 The microwave emission was
confirmed later from the excess in Planck data \citep[see][]{planck13}, whose structures coincided
nicely with the FBs. In addition, we notice that similar giant regions in the GC were found in nonthermal radio in the 1970s
\citep[seem e.g.,][]{sofu} and from X-rays observation of ROSAT \citep[see][]{Snowden1997}.

The FBs have an altitude about $7\sim 8$ kpc, which coincide spatially with the giant X-ray bubbles.
These X-ray structures were found by the eROSITA \citep[][]{predehl}.
These eROSITA Bubbles (RBs) emit in the range of thermal keV region ($0.1 \sim 2.4$ keV).
They are above the FBs with a size of about 14 kpc.
The estimated energy of the bubbles is around $10^{56}$  erg,  and the total power in X-rays is about $10^{39}$ erg s$^{-1}$.
\citet{kesh} assumed that these RBs are counterparts of the FBs
We see two envelopes (inside each other) where the inner envelope is seen as a nonthermal $\gamma$-ray/microwave emission region,
while the outer envelope is seen as the region of thermal X-rays.
It is still a question whether these structures had a common origin or were produced by unrelated events.
As suggested by \citet{kesh}, these two bubbles might be generated by two outbursts with huge release of energy from the GC
and the RBs were older than the FBs.
Below we analysed the model of FBs, and the origin of the RB envelope is beyond the scope of the present work.

For the review of the Fermi/eROSITA bubble observations and interpretation  one can find a nice review of \citet{sark} and \citet{gold26}.

Several speculations were presented in the literature to interpret processes of the huge energy release in the Galaxy,
which were needed to explain the origin of the FBs there.
The total energy, required to generate the large galactic outflow of the FBs, is assumed to be in the range up to about $10^{56}$ erg.
This energy release in the GC may be a compelling evidence for a huge energetic explosion occurred in the GC
a few ($2\sim 8$) million years ago \citep[see e.g.][]{naya,yangk}.
Alternative model of FBs was suggested \citep[see][]{crock11a,crock11b},
that their origin is due to supernova explosions in star-formation regions in the GC,
and the FBs $\gamma$-rays were produced by collisions of relativistic protons with the background gas \citep[see also][]{yang18,mou,shim}.
The observed microwave emission from the FBs was interpreted as synchrotron radiation of secondary electrons produced by {\it pp} collisions.
However, \citet{cheng15a} showed that this model was unable to match with the $\gamma$-ray and microwave data.
Nevertheless, we plan to analyse the effect of
non-linear Landau damping
for the FBs in the future \citep[see, e.g.,][]{chern22,capr}.

Alternative source of energy input of the bubbles can be sporadic stellar tidal disruption events (TDEs) of energy about $10^{53}$ erg by
the supermassive black hole at GC.
The rate of disruptions in the GC, needed for the origin of FBs, is estimated as $10^{-4}\sim 10^{-5}$ yr$^{-1}$ \citep[see, e.g.,][]{stone,goto25},
providing the average power of energy release (luminosity) from the GC into the halo of $L\sim 10^{41}$ erg s$^{-1}$ .
For the power of GC, the total energy release into the bubble is $E_{\rm tot}(t)=Lt$ where $t$ is the current time \citep[see, e.g.,][]{ko20, dog24}
and the time $t=0$ defines the initial period of star disruptions.

One of the explanation is that this $\gamma$-ray and microwave emission can be generated by CRs in the FB envelope
by the standard mechanism of shock acceleration \citep[see, e.g.,][]{cheng11,dorf12,per22,diet25}.
Tidal star disruptions near the central black hole create shock fronts inside of the bubbles that may be able to continuously accelerate CRs to high energies
if the Mach number $M\gg1$ ($M=v_{\rm sh}/c_{\rm s}$, here $v_{\rm sh}$ is the shock velocity, and $c_{\rm s}$ is the sound speed.).
This process may explain the
leptonic origin of microwave and $\gamma$-rays in the halo of FBs by a repeated star captures in the GC \citep[see e.g.][and references therein]{ko20}.
Nevertheless, some discrepancies between this model and observations need  further investigations for the shock model.

 The parameters of the FB shocks can be inferred from the intensity of soft X-rays \citep[see][]{kata13,kata15} and Oxygen lines \citep[see][]{mill16}.
For a weak shock with a low Mach number $M\sim 1.5 - 2.3$ that gave the shock envelope a velocity about $300 - 500$ km s$^{-1}$.
As it follows from the model of FB shock acceleration \citep[see][]{diet25} the estimated Mach number ($M \sim 2 - 3$)
might be too low for accelerating CRs to high enough energies if the average power is about $3\times 10^{41}$ erg s$^{-1}$.

An alternative mechanism, CR stochastic acceleration, was presented in, e.g.,
\citet{mert1,cheng15b, mert,yangk,tseng24}.
This stochastic acceleration of CRs by turbulence appears to be more plausible in comparison with the model of  shock acceleration.

\section{The model of stochastic acceleration in FBs}

The lack of evidence of a strong shock coinciding with the edge of the FB bubbles and constraints from multi-wavelength observations,
point towards stochastic acceleration of CRs by turbulence
is a likely mechanism.
Plasma instabilities, in particular Rayleigh-Taylor and Kelvin-Helmholtz instabilities \citep[see, e.g.,][]{yang12},
would generate turbulence at a shock that is convected into the bubble interior by the downstream plasma flow.

\citet{cheng14} suggested a model where CRs are accelerated in-situ directly by the background plasma of the FBs via stochastic hydromagnetic turbulence \citep[see][]{gur,dog00}.
One of the main problems of this model is that this process will overheat the plasma by accelerated particles \citep{wolfe06,east08},
because the energy transferred to accelerated particles is quickly dumped into the thermal plasma.
\citet{chern12} showed, however, that the effect of overheating depends on parameters of acceleration, and can be insignificant.

\citet{cheng15b} put forward the idea that routine tidal disruption of stars by the supermassive black hole at the GC would generated MHD-turbulence in the Galactic halo,
which propagate through the exponential halo atmosphere.
In this case CR electrons, generated by supernova remnants (SNRs) in the Galactic Disk, are re-accelerated the Galactic halo.
They concluded that the $\gamma$-ray and microwave spectra of the Fermi bubbles could be reproduced by the inverse-Compton and synchrotron emission by the accelerated electrons.
The main problem of the model is the coefficients of momentum and spatial diffusion for the stochastic acceleration in the derived kinetic equation in \citet{cheng15b}
are rather arbitrary, which are estimated somehow from observations.

\citet{mert} considered that the main parameter of the model is the mean turbulent energy density in the Galactic halo whose velocity of turbulence is
$\delta {\rm v}_{\rm turb} \approx 10$ km s$^{-1}$ and
the energy density of the turbulence is $w_{\rm turb}\approx 5 \times 10^{-3}$ eV cm$^{-3}$.
The total turbulent kinetic energy for one bubble of 4 kpc radius is $W_{\delta  {\rm v}}\sim 7 \times 10^{52}$ erg,
which is about a percent of the total kinetic energy provided by the inner Galaxy.
They included the effects of stochastic acceleration by turbulence and estimated the momentum and spatial diffusion coefficients of CR kinetic equation in the magnetized medium of the bubble.
They concluded that the $\gamma$-ray as well as microwave spectral and morphological features of the Fermi bubbles can be reproduced by the inverse-Compton and synchrotron emission
from electrons accelerated by turbulence generated in a mildly supersonic outward flowing shell.

The goal of our analysis is to derive (not estimate) the parameters of the kinetic equations for both CRs and MHD-turbulence.
The key parameters of our model come from \citet{breit22} where the spectrum of MHD-turbulence was derived from the characteristics of Rayleigh-Taylor instabilities in the FBs.
They derived the time-dependent criteria for shock promoting the interpenetrating spikes and their mixing for the cases of constant, exponential, and power-law background.

They developed two models of giant envelopes which can be conditionally accepted as FBs and RBs.
The first model was described by the Kompaneets formalism for exponential atmospheres.
They estimated the time for RT instabilities and the time for FB envelope fragmentation.
The top of the envelope can be destroyed by Rayleigh-Taylor instability \citep[see][]{baumbr13}.
The age of the FBs may be defined by the time of shell break-up.

For power-law gas distribution in the halo, \citet{ko20} derived parameters of the envelope propagation in the power $\beta$-atmosphere (for $\beta=3/2$).
In this limit the velocity of the envelope decreases for a single input of energy, or converged to a constant velocity for sequential inputs of energy.
The RT instability is not expected in this scenario.

The gas distribution of the halo is uncertain at the moment.
For instance, \citet{cordes} and \citet{biswas} suggested an exponential profile as $\rho(z)\propto e^{-z/H}$,
whereas \citet{mill95,mill16} proposed a so-called $\beta$-model profile, which takes the form $\rho(z)\propto(1+z/H)^{-3\beta}$ (e.g., $\beta = 2/3$).
Here $H$ is a scale of the gas distribution in the Galactic halo.

\citet{breit22} ruled out RT turbulence for the RB envelope of the $\beta$-gas distribution. The outer shell has not been affected by RT instabilities.
Therefore, the dynamical evolution of the bubbles suggests a maximum final age about 20 Myr.
This may imply that FBs fill only about 15 per cent (or even less) of their final size of the RBs.
The energy input from the GC region may occur not in a constant but in an episodic fashion, which is not unlikely for TDEs.

In this paper we intend to present an alternative non-linear model for CR acceleration in the FBs.
The CR spectrum is derived from the non-linear equations of MHD-hydrodynamic and kinetic propagation of CRs.
The parameters in the model are derived quantities not estimated ones.

\section{X-ray halo}

The bubble evolution was analysed for a  medium with an inhomogeneous
gas density  \citep[see][]{baumbr13,breit22,diet25}.
Below, we
define
the parameters for the Fermi bubble in the Galactic halo.

The first indication on the X-ray structure the halo was found by \citet{Snowden1997} from ROSAT, which was seen as a bulge of hot gas
similar to a cylinder with an exponential fall-off of density with height above the plane.
The cylinder has a radial extent around 5.6 kpc, and the scale height of 1.9 kpc with electron density at the base about
$0.0035$ cm$^{-3}$ and temperature about $10^6$ K.

Unlike in the disk, the gas distribution in the halo is not very reliable.
For instance, \citet{cordes,biswas} assumed that the plasma density in the Galactic halo drops exponentially with height $z$ above the Galactic plane.
Let $n_0$ be the gas number density at $z=0$ and $H$ is the density scale height.
\citet{nord92} estimated the number density of free electrons above the plane as $n_0=0.033$ cm$^{-3}$
and the characteristic scale of the electron distribution there as $H\sim 0.53 - 0.84$ kpc.
Similarly, \citet{gaen08} derived the warm plasma distribution ($\sim 10^4$ K) above the Galactic disk
($\leq 2$ kpc with the average density $\sim 0.014$ cm$^{-3}$)
from pulsar dispersion measures and H$\alpha$ diffuse emission.

From Suzaku observation \citep{naka18,Nakashima2019} a disk-like density distribution
of the hot gas ($T\sim 10^6$ K) in the halo was derived
\begin{equation}
\rho(r,z)=\rho_0\exp\left(-\frac{r}{r_0}\right)\exp\left(-\frac{z}{H}\right)\,,
\label{rhz}
\end{equation}
with $\rho_0\simeq 7\times 10^{-27}$ cm$^{-3}$, the scale height $H\simeq 1.3$ kpc, and the radial scale length $r_0\simeq 7$ kpc.

\section{ Qualitative Picture of Shock Evolution in the Galactic Halo.}

Below we develop a model of CR acceleration by MHD-fluctuations excited by the Rayleigh-Taylor (RT) instabilities in the shock envelope of FBs.

We present the analysis of shock evolution in the halo as discussed in \citet{baumbr13}.
Their results are based on the analytical solution of \citet{komp60} for an exponential atmosphere.
For a toy-model of the FBs we use a relatively free set of parameters of a qualitative halo model.
The pressure of background gas is neglected.

We  adopt the gas exponential distribution of the gas in the halo as
\begin{equation}
	\rho(z) = \rho_0 \exp\left(-\frac{z}{H} \right)\,,
\label{a1}
\end{equation}
where $\rho_0$ is the gas density at $z=0$ (the base of the Galactic halo),
$H$ is the scale height of the halo.

The analytical solution for a shock wave propagation in the exponential halo  was obtained by \citet{komp60}
which is presented below in the review of \citet{kogan}.
The shock front is described as
\begin{equation}
 \left(\frac{\partial r}{\partial y}\right)^2
  = \exp\left(\frac{z}{H} \right) \left[\left(\frac{\partial r}{\partial z}\right)^2+1\right]\,,
\label{eq:kompaneets}
\end{equation}
where $y$ is a transformed time variable
\begin{equation}
  y(t)=\int\limits_0^t \sqrt{\frac{(\gamma_{\rm g}^2-1)}{2}\frac{E(t)}{\rho_0V(t)}}\,dt\,,
\label{eq:timevarible}
\end{equation}
where $\gamma_{\rm g}=5/3$ is the adiabatic index,
$E(t)$ is the total energy release (if $L$ is the luminosity of the GC and $t$ is the time since the injection, then $E(t)=Lt$).
$V(t)$ is the current volume of the expanding bubble,
\begin{equation}
  V(t)= 2\pi\int\limits_{0}^{z_u}r^2(z,t)\,dz\,.
\label{eq:volume}
\end{equation}
Here $r(z,t)$ is the radius of the envelope in cylindrical coordinates, and $z_u(t)$ is the height of the top of the bubble.
Factor of $2$ appears because of two bubbles, above and below the plane.

For convenience, we introduce dimensionless variables,
\begin{equation}
\tilde{z} = \frac{z}{H}\,,  \quad
\tilde{r} = \frac{r}{H}\,,  \quad
\tilde{V} = \frac{V}{H^3}\,,  \quad
\tilde{t} = \frac{t}{t_0}\,,  \quad
\tilde{y} = \frac{y}{H}\,,  \quad
\tilde{m} = \frac{m}{\rho_0 H^3}\,,
\label{eq:norm1}
\end{equation}
where
\begin{equation}
t_0^3 = \frac{2}{\left(\gamma_{\rm g}^2 -1\right)} \cdot \frac{\rho_0 H^5}{L}\,.
\label{tilde}
\end{equation}

The self-similarity of the solution implies that its dimensionless shape does not depend on the values of
the model parameters $H$, $\rho_0$ and ~$L$.
Thus, calculations for the same dimensionless variables can be performed for different
models by simply changing the scales according Eq. (\ref{eq:norm1}).

For these variables we can derive a solution in the dimensionless form.
Returning from the dimensionless solution to dimensional variables, we will estimate the parameters of FBs from observational data.

In dimensionless variables,
\begin{equation}
\tilde{V}(\tilde{y}) = 2\pi \int\limits_{0}^{\tilde{z}_u} \tilde{r}^2(\tilde{y},\tilde{z})\,d\tilde{z}\,,
\end{equation}
where the position of the shock front is derived from Eq. (\ref{eq:kompaneets})
\begin{equation}
\tilde{r}(\tilde{y},\tilde{z}) = 2 \cos^{-1} \left[\frac{1}{2} e^{\tilde{z}/2} \left(1-\frac{\tilde{y}^2}{4} + e^{-\tilde{z}}\right)\right] \,.
\end{equation}
The $z$-coordinates of the bubble top~$z_u$ and of the maximum radius~$z_0$ are:
\begin{equation}
\tilde{z}_u = -2 \ln \left(1-\frac{\tilde{y}}{2}\right)\,,  \qquad
\tilde{z}_0 = -\ln \left(1-\frac{\tilde{y}^2}{4}\right)\,.
\label{zu}
\end{equation}
One can rewrite \eqref{eq:timevarible} in dimensionless variables and get
\begin{equation}
\left(\frac{d\tilde{y}}{d\tilde{t}}\right)^2 \cdot \tilde{V}(\tilde{y}) = \tilde{t} \,,
\label{tt1}
\end{equation}
and define the time $\tilde{t}$ as a function of the variable $\tilde{y}$,
\begin{equation}
\tilde{t}(\tilde{y})=\left(\frac{3}{2}\int\limits_0^{\tilde{y}} \sqrt{\tilde{V}(\tilde{y})}d\tilde{y}\right)^{2/3}\,.
\label{tt2}
\end{equation}
\begin{figure}
\includegraphics[width=0.8\columnwidth]{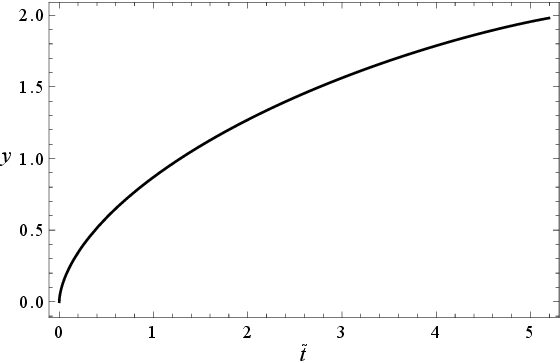}
\caption{Dependence of the transformed time
$\tilde{y}$
on the dimensionless time $\tilde{t}$.
}
\label{fig:y_t}
\end{figure}

\begin{figure*}[t]
\centering
\includegraphics[width=0.45\columnwidth]
{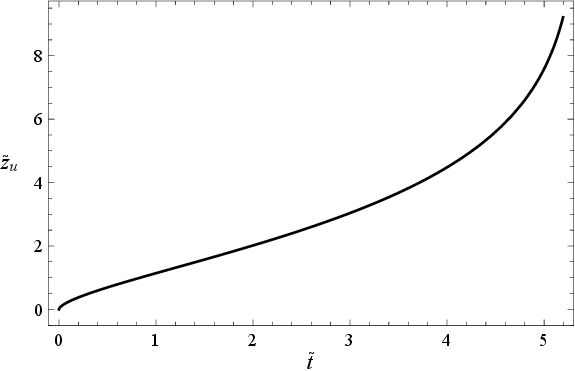}
\includegraphics[width=0.45\columnwidth]{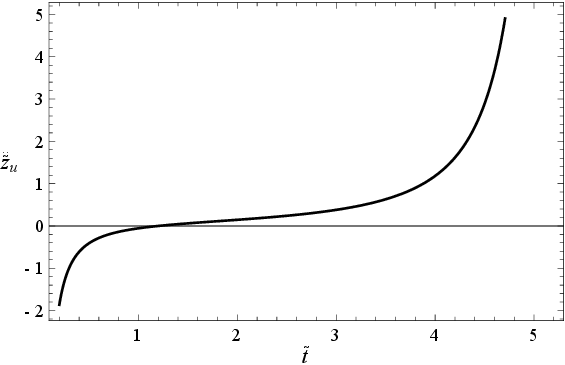}
\caption{{\it
Left}, Position of the top of the bubble as function of the dimensionless time $\tilde{t}$. {\it Right},
Acceleration of the FB envelope at the top the shell.}
\label{fig:z_t}
\end{figure*}

The function
$\tilde{y}(\tilde{t})$
is presented in Fig.~\ref{fig:y_t}.
For the exponential gas distribution (\ref{a1}), the transformed time cannot exceed the value of
$\tilde{y}(\tilde{t})<2$.
The corresponding position of the bubble top
$\tilde{z}_u(\tilde{t})$
is presented in Fig.~\ref{fig:z_t}a.

The velocity of the envelope decreases at the initial stage (Sedov solution).
It changes to acceleration at time $\tilde{t}_\eta$ if its velocity exceeds the sound speed \citep[see, e.g.,][]{dog24}.
The top of the bubble begins to accelerate when $\tilde{t}>\tilde{t}_\eta\sim 1.202$ (see Fig.~\ref{fig:z_t}b).

The velocity and acceleration of the top of the bubble were presented in \citet{baumbr13}
\begin{equation}
\dot{\tilde{z}}_u(\tilde{t}) = \frac{d\tilde{z}_u}{d\tilde{y}}\frac{d\tilde{y}}{d\tilde{t}} =
\frac{1}{\left(1-\tilde{y}/2\right)} \cdot \frac{d\tilde{z}_u}{d\tilde{y}}\,,
\end{equation}
and
\begin{equation}
\ddot{\tilde{z}}_u(\tilde{t}) = \frac{d^2 \tilde{z}_u}{d\tilde{t}^2} =
\frac{d \dot{\tilde{z}}_u}{d\tilde{t}} =
\frac{d \dot{\tilde{z}}(\tilde{y})}{d\tilde{y}} \frac{d\tilde{y}}{d\tilde{t}}\,.
\label{zzt}
\end{equation}

The acceleration at the top of the bubble, ${\ddot z}_u(y)$,
may produce the RT instability if the envelope acceleration is high enough.
Inside the envelope the RT instabilities are excited between the dense shell $\rho_{\rm sh}$ and the hot interior $\rho_{\rm in}$
\citep[see]
[
 and for the effect of magnetic fields for the RT instabilities see \citealt{bian}]
{baumbr13,breit22}
when the envelope is accelerated into the exponential halo (see Eq. (\ref{zzt})).
\begin{equation}
\tau_{\rm RT}(z_u(y))=\sqrt{\frac{d(y)}{2\pi {\ddot z}_u({y})}
\frac{[\rho_{\rm sh}(y)+\rho_{\rm in}(y)]}{[\rho_{\rm sh}(y)-\rho_{\rm in}(y)]}\,}\,.
\label{ttrt}
\end{equation}
The RT instability plays a key role for the envelope evolution.

\section{ Power of turbulence in the FB envelope}

The gas of the bubble is accumulated by the envelope into the thin shell (shock) of thickness $d(t)$.
The total gas inside the envelope is $m(t)$ at the current time
\begin{equation}
m(t) = 2\pi \int_0^{z_u}\rho_0 e^{-z/H} \cdot r^2 \cdot \, dz\,.
\end{equation}

Following \citet{breit22}, we define the shell thickness $d(t)$ as
\begin{equation}
m(t) = d(t) \times \int_{\rm surfase} \rho_{\rm in} dS\,.
\end{equation}

As an example, in the limit of strong shock ($M\gg 1$),
$\rho_{\rm in}(z) = 4 \rho_{\rm out}(z)$ is defined from  \citet{breit22} (see there for details), then the shell thickness, $d(t)$, is
\begin{equation}
d(t) = m(t)\left[8\pi\rho_0 \int_0^{z_u} e^{-z/H}\cdot r
\sqrt{1+\left(\frac{\partial r}{\partial z}\right)^2\,} \, dz\right]^{-1}\,.
\end{equation}

During the nonlinear regime  their temporal evolution of the amplitude of RT instability, $\eta$, is obtained by
numerically solving of ordinary differential equation for the RT instabilities \citep[see][]{breit22},
\begin{equation}
{\dot\eta}(\tilde{y})=2\sqrt{\alpha {\ddot z}_u(\tilde{y})\eta(\tilde{y})\,}\,,
\label{zdot}
\end{equation}
whose numerical parameter $\alpha$ is within the range $0.02 -0.1$ depending on models.
We take $\alpha=0.06$ in the following calculations from \citet{breit22}.
From Eq.~(\ref{zdot}) we get
\begin{equation}
\eta(\tilde{t}) = \alpha \left( \int_{\tilde{t}_{\rm start}}^{\tilde{t}} \sqrt{{\ddot z}_u\,}\, d\tilde{t} \right)^2 \,.
\label{dot}
\end{equation}

\begin{figure*}[t]
\centering
\includegraphics[width=0.45\columnwidth]
{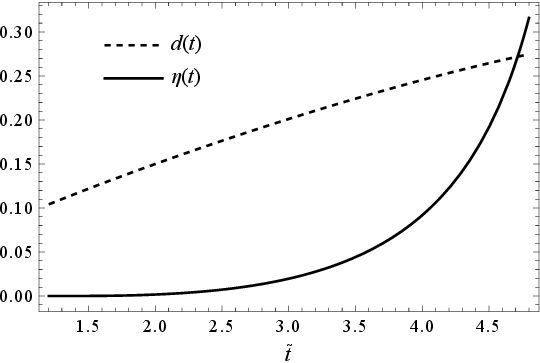}
\includegraphics[width=0.45\columnwidth]{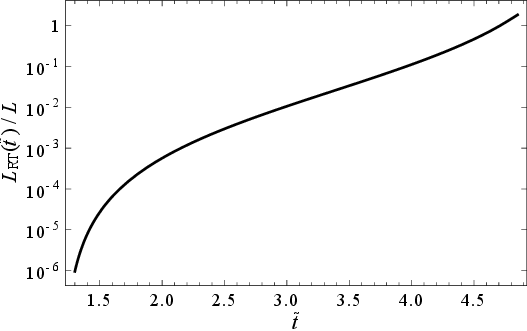}
\caption{{\it
Left}.The amplitude of the Rayleigh-Taylor instabilities $\eta(t)$ and the thickness of the bubble $d(t)$. {\it Right}.
The ratio of the luminosity of RT turbulence to the total luminosity at the GC derived from $L_{\rm RT}(\tilde{t})/L$.}
\label{fig:d_t}
\end{figure*}

The functions $d(t)$ and $\eta(t)$ are shown in Fig.~\ref{fig:d_t}a.
At the time $t_{\rm br}$ the RT amplitude $\eta(t)$ is of the order of $d(t)$,
and the shock shell is completely
disrupted.

From numerical calculations we get the values at this time
\begin{eqnarray}
&&\tilde{t}_{\rm br}=4.715\,, \qquad \tilde{y}(\tilde{t}_{\rm br})=1.913\,,
\qquad \tilde{z}_u(\tilde{t}_{\rm br}) = 6.27,\nonumber\\
&&\tilde{\eta}(\tilde{t}_{\rm br}) = 0.272\,,
\qquad \dot{\tilde{\eta}}(\tilde{t}_{\rm br}) = 0.467\,.
\label{para}
\end{eqnarray}

The RT instability is excited by a hydrodynamic turbulence, and the time of turbulence excitation can be estimated as
\begin{equation}
\tilde{t}_{RT} \approx\frac{\tilde\eta(\tilde{t}_{\rm br})}{\dot{\tilde{\eta}}(\tilde{t}_{\rm br})} \approx 0.58\,,
\end{equation}
 where $\tilde{t}_{\rm RT}$ is
the characteristic time scale of RT instabilities, $\tilde{t}_{br}$ marks the time of disruption.

If we consider the RT turbulence in the shell has a largest length scale of $l_{\rm turb}=\eta$
and velocity at this scale is of the order of $v_{\rm turb}=\dot{\eta}$ (see Eq.~(\ref{zdot})), then
the spatial density of power, pumping by the hydrodynamic turbulence, is
\begin{equation}
\varepsilon_0(t) = \rho \, \frac{v_{\rm turb}^3}{l_{\rm turb}}\sim 4\rho(z_u(t))\frac{\dot \eta(t)^3}{\eta(t)}\,,
\label{vareps}
\end{equation}
where $\rho(z_u)$ is given by Eq. (\ref{a1}), $v_{\rm turb}=v_0$ and  $l_{\rm turb}=2\pi/k_0$ (see below).

We estimate the mass of turbulent gas as the mass of the bubble above the height with maximum radius,
\begin{equation}
m_{\rm turb} = 2\pi \int_{z_0}^{z_u}\rho_0 e^{-z/H} \cdot r^2 \cdot \, dz\,.
\end{equation}
Then, the total power of RT turbulence in the shell in can be estimated as
\begin{equation}
L_{\rm RT}(t) \simeq m_{\rm turb} \, \frac{{\dot{\eta}}^3}{\eta}\,.
\end{equation}
With Eqs.~(\ref{eq:norm1}), (\ref{tilde}), \& (\ref{zdot}), we have
\begin{equation}
\frac{L_{\rm RT}(\tilde{t})}{L} =  4 ({\gamma_{\rm g}}^2 -1) \alpha^{3/2}\, \tilde{m}_{\rm turb} \, {\ddot{\tilde{z}}_u}^{3/2}\, \tilde{\eta}^{1/2}\,.
\end{equation}

The function $L_{\rm RT}(t)/L$ for $\alpha=0.06$ is shown in Fig.~\ref{fig:d_t}b.
At the final stage $\tilde{t}_{\rm br}$ the RT power can be of the same of order as the power, $L$, of TDEs  in the GC,
$L_{\rm RT}(\tilde{t}_{\rm br}) \leq  L$.

The result can be compared with the two-dimensional numerical simulations of \citet{yangli} for SNR RX J1713.7-3946 blast wave.
The blast wave is interacting with a turbulent plasma background whose the magnetic field is amplified
by RT instabilities at the contact discontinuity of the shock flows.
In their paper, electrons are accelerated via stochastic scattering with magnetized turbulent plasma instead of by diffusive shock acceleration,
because most of the magnetic field is generated via the RT instability near the contact
discontinuity and turbulent motion in the shock downstream.

\section{Source of MHD-waves excited by hydrodynamic turbulence in the envelope of Fermi bubbles}

In this section we present the mechanism of  MHD-wave excitation by turbulence.
\citet[][]{light} suggested a direct source of transferring hydrodynamic turbulent energy to acoustic waves.
Turbulent eddies release kinetic energy to surrounding waves.
Most of this energy is returned to the ambient medium, but a small fraction transforms into propagating acoustic waves \citep[][]{batch70}.

\citet{kuls55} (and later \citet{parker,kato}) developed this theory for incompressible MHD-waves
in the approximation of ``strong'' magnetic fields when the velocity of hydrodynamic turbulence $v_{l}$,
pumping on the scale $l$, is smaller than the Alfv\'en velocity $v_A$, i.e.,
\begin{equation}
M_A=\frac{\langle v_{l}^2\rangle^{1/2}}{v_A} <1\,.
\end{equation}
where $v_A=B/\sqrt{4\pi \rho}$.  Contrary to shock acceleration, turbulent resonant acceleration does not require strong shocks.

For small Mach numbers ($M_A< 1$, strong magnetic fields) a fraction of the energy emits in the form of Alf\'ven waves.
The hydrodynamic turbulence on a scale $k$ with turbulence velocity $v$ will radiate Alfv\'en waves at the wavenumber $\bar {k}$
\begin{equation}
\bar{k}=\left[\frac{v}{v_A}\right]k\,,
\label{61}
\end{equation}
which follows from the condition that RT turbulence and resonant
Alfv\'en waves oscillate at the same frequency and from the corresponding dispersion relations.

In the limit of ``strong'' magnetic field, $v_A\gg v_l$,
the Kolmogorov-Obukhov spectrum of hydrodynamic turbulence \citep[see][]{kolm, obukh,batch70,zhou} is developed  in
the inertial interval of wave number $k$ for  the input energy flux $\varepsilon_0(t)$ presented by Eq. (\ref{vareps}),
\begin{equation}
\varepsilon_0\sim W(k)^{3/2}k^{5/2}={\rm constant}\,,
\label{11a}
\end{equation}
and
\begin{equation}
W(k)=C_k\varepsilon_0^{2/3}k^{-5/3}\,,
\label{kolm}
\end{equation}
where $W(k)$ is a power spectrum for fluid turbulence depended upon wave number $k$,
and $C_k$ is the Kolmogoroff constant, $C_k\sim 1$ \citep[see e.g., $C_k\sim 0.5$][]{sree}.

For these processes the total power of
hydrodynamic
turbulence per unit volume, going
into the Alfv\'en mode, is \citep[see][]{hendr,eilek84,fuji}
\begin{equation}
P_A(t)=\eta_A\left(\frac{E(t)}{\rho v_A^2}\right)^2\rho v_A^3k_0\,,
\label{panew}
\end{equation}
where
$E(t)=\rho {v_0}(t)^2$
is the energy density of hydrodynamic turbulence and the coefficient $\eta_A\sim 1$.

This equation can be converged to \citet{kato}
\begin{equation}
P_A(t)\approx \varepsilon_0(t)\,\frac{v_{0}(t)}{v_A(t)}\,,
\label{sato}
\end{equation}
for $M_A < 1$.

The differential energy power into Alfv\'en waves can be presented as
\begin{equation}
I_A(\bar{k},t)=I_0(\bar{k}/k_0)^{-3/2}\,,
\label{ia}
\end{equation}
where
\begin{equation}
I_0\approx \frac{1}{2}\rho v_0^3\left(\frac{E(t)}{\rho v_A^2}\right)^{3/4}\,.
\label{eq:I_0}
\end{equation}

The total power-law of MHD-waves $L_A$ can be excited by the Lighthill mechanism of hydrodynamics
\begin{equation}
L_A\sim V_{\rm FB}P_A(t)\leq 10^{40}\mbox{erg s$^{-1}$}\,.
\end{equation}
Here $V_{\rm FB}$ is the volume of MHD-wave generation in the FB.

For the parameters of the top of the FB at the time of disruption $\tilde{t}_{\rm br}=4.7$ (see Eq.~(\ref{para})),
$B=2$ $\mu$G and $L_{\rm RT}(\tilde{t}_{\rm br})=10^{41}$ erg s$^{-1}$ (see Fig. \ref{fig:d_t}b),
we estimate the Alfv\'en velocity  $v_A=1.5\times 10^8$ cm s$^{-1}$, the velocity of RT turbulence is
$\dot{\eta}\sim 4\times 10^7$ cm s$^{-1}$, and roughly from Eq.~(\ref{sato})
$L_A\sim 0.25\times 10^{41}$ erg s$^{-1}$.

\section{Generation of MHD-Waves by hydrodynamic turbulence}

The time evolution of incompressible MHD-turbulence is described by a nonlinear equation of diffusion of wave cascading
of the spectrum of wave density $W_k(\bar{k})$ \citep[see e.g.][]{gold95,gold97,mill95,cho02}
\begin{equation}
\frac{\partial W_k}{\partial t}=\frac{\partial}{\partial \bar{k}}\left(D_{kk}\frac{\partial W_k}{\partial \bar{k}}\right)-\Gamma(\bar{k})W_k+S\,.
\label{eq:spectrum_eq}
\end{equation}

Here the diffusion coefficient in the space $\bar{k}$ is $D_{kk}=\bar{k}^2/\tau_s$,
$\tau_s$ is the spectral energy transfer time and $\bar{k}$ is the wave number,
$\Gamma(\bar{k})W_k$ describes the wave damping, and $S$ is an external source of waves $S=I_A(\bar{k})$ as presented by Eq. (\ref {ia}).

The diffusion coefficient $D_{kk}$ is defined by a nonlinear system of MHD-equations for velocity $v_\lambda$
and a magnetic field ${\bf B}$  of oscillations \citep[see, e.g.,][]{srid}.
The magnetic field consists of a uniform background field and fluctuating fields,
${\bf B} = {\bf B}_0 + {\bf b}$.

The perturbed velocity and magnetic fields provided by waves traveling antiparallel/parallel to $B_0$.
The result of collisions between oppositely directed waves provides the cascade of MHD-turbulence energy.

The strength of perturbation in amplitude of velocity and magnetic fields can be characterized by
a nonlinear parameter of the frequency of fluctuations to the Alfv\'en frequency
\begin{equation}
\zeta_\lambda\sim \frac{k_\perp v_\lambda}{k_z v_A}\,,
\end{equation}
where $k_\perp$ and $k_z$ are the perpendicular and parallel components of the wave vector \citep[see][]{gold95}.

The turbulent energy $\varepsilon\sim {v_\lambda}^2/t_c$ is transported where $t_c$ is the cascade time
\begin{equation}
t_c\sim N(k_zv_A)^{-1}\,,
\end{equation}
and $N$ is the number of collisions needed for the packet to lose the memory of the initial state
\begin{equation}
N\sim {\zeta_\lambda}^{-4}=\left(\frac{k_zv_A}{k_\perp v_\lambda}\right)^4\,.
\end{equation}

For the wavelength $\lambda$ the eddy-turnover time is $\lambda/v_\lambda$ and the Alfv\'en crossing time is $\lambda/v_A$.
In the Kolmogoroff limit of strong turbulence $k_\perp v_\lambda \sim k_z v_A$,
when fluctuations of comparable size interact in one overtime, $\tau_s\sim \tau_A\sim \tau_\lambda$
(here $\tau_s$ is the spectral energy transfer time).
The wave cascading forms the the Alfv\'en wave turbulence $W_k(\bar{k})\propto \bar{k}^{-5/3}$.
The velocity fluctuations $\delta v$ is related to the rms wave magnetic field $\delta B$ by $\delta v^2/{v_A}^2\sim \delta B^2/{B_0}^2$
with $\delta B^2/8\pi\approx W_k(\bar{k})\bar{k}/2$, and the diffusion coefficient for the Kolmogoroff spectrum is
\citep[e.g.,][]{mill95}
\begin{equation}
D_{kk}  \simeq v_A \bar{k}^{7/2}\left[\frac{W_k(\bar{k},t)}{2W_B}\right]^{1/2}\,,
\end{equation}
where $W_B=B_0^2/8\pi$.

In the opposite case, $k_\perp v_\lambda \gg k_z v_A$, which is beyond our current model.
The energy transfer is defined by nonlinear interaction of  nearby MHD waves. There are
 several controversial conclusions of this model in the literature,
e.g. the spectum $W(\bar{k})\propto \bar{k}^{-3/2}$ from \citet{irosh64,nara,mill95},
or the spectrum $W(\bar{k})\propto \bar{k}^{-4/3} $ from, e.g., \citet{srid,gold95,gold97,cho02,cho03}.

\section{The spectrum of MHD-Waves and diffusion coefficients in the FB shell}
 The spectrum of MHD-waves, $W_k(\bar k)$, can be derived from the kinetic equation (\ref{eq:spectrum_eq}).
 Below $\Gamma_{\rm CR}$ is the rate of wave absorption by the friction between CRs and MHD waves, which can be found from \citet{ber90}
\begin{equation}
\Gamma_{\rm CR}(\bar{k})=\frac{\pi Z^2
e^2V_A^2}{2\bar{k}c^2}\int\limits_{p_{\rm res}(\bar{k})}^\infty\frac{dp}{p}F(p)\,,
\end{equation}

As it was shown in \citet{cheng14} this effect of absorption is essential at low CR energies which were estimated from the balance between absorption
and wave cascade terms.
Therefore,  we presented a truncated diffusion equation of (\ref{eq:spectrum_eq}),
if the effect of CR damping is insignificant and $2\Gamma_{\rm CR}W_k$ can be ignored.
We will study the effect of
non-linear Landau damping
\citep[][]{chern22,capr} on the system in future work.

Then the spectrum $W(\bar{k},t)$ is simply derived from the balance of wave cascading and MHD-excitation by the hydrodynamic turbulence,
\begin{equation}
-\,\frac{v_A}{\sqrt{2W_B}}\frac{\partial}{\partial\bar{k}}\left[\bar{k}^{7/2}W_k^{1/2}\frac{\partial W_k(\bar{k},t)}{\partial\bar{k}}\right]
\simeq
I_0\left(\frac{\bar{k}}{k_{0}}\right)^{-3/2}\,,
\label{111}
\end{equation}
where $k_{0}$ is the maximum scale of RT turbulence (see Eq. (\ref{vareps})).

%

Then the function of $W_k(\bar{k},t)$ is
\begin{equation}
W_k(\bar{k},t)=\bar{k}^{-5/3}\left[\frac{\sqrt{2W_B}}{v_A}I_0k_0^{3/2}\left(\bar{k}_0^{-1/2}-\bar{k}^{-1/2}\right) \right]^{2/3}\,,
\label{fre1}
\end{equation}
where $\bar{k}_0$ from Eq. (\ref{61}).
It is conveniently to transform Eq. (\ref{fre1}) into the form
\begin{equation}
W_k(\bar{k},t)=A\,\left(\frac{\bar{k}}{\bar{k}_0}\right)^{-2}
\left[\left(\frac{\bar{k}}{\bar{k}_0}\right)^{1/2}-1\right]^{2/3}\,,
\label{fre}
\end{equation}
where
\begin{equation}
A=2^{-2/3}\rho(t)\,v_0^2\,\frac{v_0}{v_A}\,\frac{k_0}{\bar{k}_0^{2}}\,.
\label{factorAprime}
\end{equation}

\begin{figure}
\includegraphics[width=0.8\columnwidth]{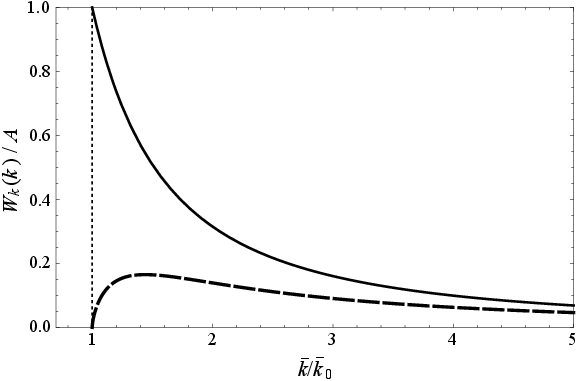}
\caption
{
The thick dashed line shows the normalized spectrum, $W_k(\bar{k},t)/A$ (where $A$ is given by Eq.~(\ref{factorAprime})).
The wave-number $\bar{k}_0$ is the scale of Alfv\'en waves, pumped by the RT hydrodynamic turbulence.
The solid line shows the power-law approximation of the spectrum $\sim (\bar{k}/\bar{k}_0)^{-5/3}$.
}
\label{lt}
\end{figure}

The expected spectrum $W_k(\bar{k},t)$ is shown in Fig. \ref{lt}. In the figure, the thick dashed line is derived from Eq. (\ref{fre}).
The MHD-waves  are excited by the RT turbulence at the injection  wavenumber $\bar{k}_0\leq k$. For comparison,
the solid line is the power-law approximation (Kolmogoff spectrum).

The rate of stochastic acceleration (momentum diffusion $D_p)$ in the kinetic equation for CRs \citep[see, e.g.,][]{dog24})
can be derived from  the spectrum $W_k$,
\begin{equation}
D_p(p,t)= \frac{4\pi^2v_{\rm A}^2 \bar{k}}
{c}\cdot\frac{\bar{k}W_k(\bar{k})}{B^2} p^2 \,.
\label{mdif}
\end{equation}

Below we will derive simple estimates for the maximum energy of electrons in the FB envelope.

\section{Acceleration of CR electrons}

In the framework of standard model of  propagation CRs are ejected by  sources in the Galactic disk,
and confined in an extended halo before escaping the Galaxy
\citep[see, e.g.,][and the numerical code GALPROP \citealt{mosk98,vladim}]{ginz64,ber90}.
CR electrons of more than several GeV are unable to reach high altitude above the Galactic plane because of energy losses.

 \citet{cheng15b} developed a phenomenological model of stochastic acceleration for the model of CR origin in the FBs for appropriated parameters.
 The kinetic equation for
the distribution function of electrons, $f(r,z,p)$, in this case is
\begin{eqnarray}
&&-\nabla \cdot\left[\kappa(r,z,p)\nabla f -u(r,z)f\right]+\nonumber\\
&&\frac{1}{p^2}\frac{\partial}{\partial p}p^2\left[
\left(\frac{dp}{dt}-\frac{\nabla\cdot {\bf u}}{3}p\right)f -
D_p(r,z,p)\frac{\partial f}{\partial p}\right] =
Q(p,r)\delta(z) \,,
\label{eq_nu}\end{eqnarray}
where $u$ is the
velocity of the Galactic wind, $\kappa$ and $D_p$  are the spatial
and momentum (stochastic acceleration) diffusion coefficients,
$(dp/dt)$ describes the rate of electron energy losses, and $Q$ describes CR sources \citep[for details, see][]{cheng15b}.

The solution of this equation was obtained in the analytical form and this re-acceleration model reproduced reasonably well
the spectra of $\gamma$-ray and microwave emission from the FB as well as the sharp edge at the outer boundary of the FB
\citep[see Figs. 5 and 8 in][]{cheng15b}. Their calculations showed that
the re-acceleration mechanism is effective for the production of electrons with energies below $10^{12}$eV.

Below we will present a revised alternative model acceleration of CR electrons by the RT turbulence
and MHD-wave excitation for estimations of maximum energies of electrons.

The maximum energy of accelerated electrons can be estimated from the equity of their acceleration time and their loss time.
According to Eq.~(\ref{mdif}), the acceleration time can be estimated as
\begin{equation}
  \tau_{\rm acc} = \frac{p^2}{4D(p)}
  \sim \frac{2^{2/3}c}{\pi}\,\frac{\eta^{2/3}}{{\dot \eta(t)}^2}
  \left(\frac{\dot \eta(t)}{v_A(t)}\right)^{1/3}\left(\frac{pc}{eB}\right)^{1/3}\,.
  \label{eq:acc_time_expanded}
\end{equation}
As $\eta = H\tilde{\eta}$ and $\dot{\eta} = H\dot{\tilde{\eta}}/t_0$,
\begin{equation}
    \tau_{\rm acc} = \frac{2^{2/3}c}{\pi}\,\frac{\tilde{\eta}^{2/3}}{{\dot{\tilde{\eta}}(t)}^2}
    \left(\frac{\dot \eta(t)}{v_A(t)}\right)^{1/3} \frac{r_L^{1/3}t_0^2}{H^{4/3}} \,,
\end{equation}
where $r_L$ is the Larmor radius of electron.

\begin{figure}
\includegraphics[width=0.8\columnwidth]{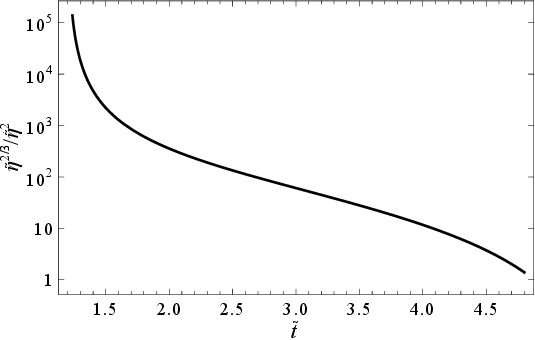}
\caption{
$\tilde{\eta}^{2/3}/\dot{\tilde{\eta}}^2$ as function of dimensionless time $\tilde{t}$.
}
\label{fig:combination}
\end{figure}

The acceleration time depends on the combination $\eta^{2/3}/\dot{\eta}^2$ at the position of the top of the shell,
which decreases with the evolution of the bubble shell.
Therefore the lowest value of $\tau_{\rm acc}$ can be estimated from the values $\tilde \eta$ and
 $\dot{\tilde \eta}$ at the time equal to $t_{\rm br}$ (see Fig.~\ref{fig:combination}).

Since $M_A \approx 0.4$ and ${\tilde{\eta}^{2/3}}/{\dot{\tilde{\eta}}(t)^2} \approx 2$, the acceleration time is
\begin{equation}
	\tau_{\rm acc} \approx 3\times 10^{11} \times \left(\frac{pc}{1~\mbox{GeV}}\right)^{1/3} \ {\rm sec}\,.
\end{equation}

The energy losses of high energy electrons $dE/dt$ are defined by the synchrotron and the inverse Compton emission \citep[see, e.g.,][]{ginz79}
\begin{equation}
\frac{dE}{dt}_{{\rm sync}+{\rm IC}}=\left(\frac{E}{m_ec^2}\right)^2\left[\frac{e^4B^2}{6\pi c^3m_e^2}+\frac{4}{3}c \sigma_Tw_{\rm ph}\right]\,,
\end{equation}
where $\sigma_T$ is the Thomson cross-section,
and $w_{\rm ph}$ is the energy density of the background photons in the Galactic halo.

The characteristic time of losses is
\begin{equation}
    \tau_{\rm loss} \sim  \frac{m_e c^2}{E}\, \frac{m_ec}{\sigma_T w_{\rm ph}}\,.
\end{equation}

Then the maximum energy of accelerated electrons $E_{\rm max}$ can be estimated from the balance between acceleration and energy losses,
\begin{equation}
E_{\rm max}\sim m_ec^2\left[\frac{2^{1/3}\pi m_e}{3\sigma_Tw_{\rm ph}}\, \frac{\dot{\eta}^2}{\eta^{2/3}}
\left(\frac{v_A(t)}{\dot{\eta}}\right)^{1/3}\left(\frac{\Omega}{c}\right)^{1/3}\right]^{3/4}\,,
\end{equation}
where $w_{\rm ph}$ is about 0.5 eV cm$^{-3}$ at an altitude $\sim 5~to ~8$ kpc above the Galactic plane and $\Omega$ is the electron gyrofrequency.

The dimension parameters of FB: $H=1.3$~kpc, $\rho_0 = 7\times 10^{-27}$ g cm$^{-3}$, $L=10^{41}$ erg s$^{-1}$, $\gamma_{\rm g} = \frac{5}{3}$, and
Eq.~(\ref{tilde}) gives $t_0 = 1.37$~Myr. From Eq. (\ref{para}), ${\eta}={\tilde{\eta}} H \approx 10^{21}$ cm,
${\dot\eta}=\dot{\tilde{\eta}} H/t_0 \approx 4.4\times 10^{7}$ cm s$^{-1}$,
and with $v_A\approx 1.5\times 10^{8}$ cm s$^{-1}$ (see end of Section 5),
we obtain approximately
\begin{equation}
E_{\rm max}\sim  1\ {\rm TeV}\,.
\end{equation}

This value is higher than  required to explain the $\gamma$-ray emission from FBs \citep{su10}.
Thus, the RT turbulence is a possible mechanism for acceleration of electrons responsible for the $\gamma$-rays from FB.
Also the characteristic time of acceleration is much shorter than the FB formation time, $t_{br}=6.5$~Myr,
the expansion of the FB shell should not significantly affect the accelerated electrons.

The acceleration time described by the Eq. (\ref{eq:acc_time_expanded}) at energies of about
$E\sim 0.1 - 1$ TeV is about $\tau_{\rm acc} \sim 1 - 3\times 10^{12}$ s.
This value is shorter than the acceleration time equal to $\tau_{\rm sim} = 5\times 10^{13}$ s used in our previous simulations of re-acceleration of electrons
in the Fermi bubbles \citep{cheng15b}, where we were able to reproduce both $\gamma$-ray and microwave emission spectra as well as spatial distribution of the
$\gamma$-ray emission. The thickness of acceleration zone, $\eta \approx 0.3$ kpc is also slightly larger than the value of $60$ pc used in \citet{cheng15b}.
Therefore with slight adjustments of the model parameters (like taking into account the spatial dependence of the acceleration time) we expect that the results from
\citet{cheng15b} can be reproduced.

In this study, however, we do not treat the acceleration parameter as ``external'' parameters.
Instead, we calculate their values consistently from the properties of plasma distribution in the Galactic halo and the total energy input from the GC.

\section{Conclusion}

In this work we present a hydrodynamic model of cosmic ray acceleration.
It describes the energy evolution from the initial energy release in the Galactic Centre (GC) to energetic electrons and
in the form of electron nonthermal emission far above the Galactic plane.
This model may explain the observed  microwave and $\gamma$-ray emission from the Galactic halo.

The conclusion of the paper can be itemized as follows:
\begin{itemize}
\item{} The energy input is defined by stellar tidal disruption events (TDEs) by the supermassive black hole at GC
with an energy release about $10^{53}$ erg every  $10^{4} \sim 10^{5}$ yrs.
The rate of energy release by the TDEs in the GC is estimated to be of the order of
$1 \sim 3\times 10^{41}$ erg s$^{-1}$.
\item{} This energy release in the the GC provides a shock penetrating into the halo,
and  that forms an accelerating envelope of several kpc in size.
\item{} At the late stage of the envelope, it propagates with supersonic velocities.
Inside the envelope the Rayleigh-Taylor (RT) instabilities
are excited between the dense shell  and the hot interior.
At the initial stage, the RT turbulence occupies a small part of the envelope
but as time proceeds, the RT turbulence fills the whole volume of FB envelope
and the RT luminosity could reach the value of $L_{\rm RT}\le 10^{41}$ erg s$^{-1}$.

\item{}

From the intensity of soft X-rays and Oxygen lines \citep[][]{kata13,kata15,mill16}, the parameters of the FB shock can be estimated.
The deduced shock is a weak one with a low Mach number $M\sim 1.5 - 2.3$.
from the model of FB shock acceleration \citep[see][]{diet25}, a low Mach number shock ($M \sim 2 - 3$) might not be able to accelerate CRs
to high enough energies if the average power is about $3\times 10^{41}$ erg s$^{-1}$.
Unlike the shock model of CR acceleration, the RT model of in-situ turbulent resonant acceleration in the FBs does not require strong shock fronts.
In this work, we derived the spectrum of RT instabilities and determine
the spectra of kinetic equations for MHD-fluctuations needed for the acceleration of CRs.

\item{} Hydrodynamic RT turbulence of the Kolmogoroff spectrum excites resonant Alfv\'en waves with the same frequency.
The power of RT turbulence is transformed partly into the Alfv\'en waves,
and at the late stage the luminosity of Alfv\'en waves can reach $P_A\sim L_{\rm RT}(v_l/v_A)$.
Here $v_l<v_A$ for strong magnetic fields.
\item{} The time evolution of MHD-turbulence $W(\bar{k},t)$ is described by a
nonlinear diffusion equation of wave cascading and injection by external sources in the form RT fluctuations.
\item{}The momentum diffusion of charged particles can be estimated from the spectrum of Alfv\'en waves,
and a maximum energy about 1 TeV of the accelerated electrons can be estimated from the balance between
the rate of stochastic acceleration (momentum diffusion) and the energy losses.
\end{itemize}

\section*{Acknowledgments}
 The authors are grateful to Dieter Breitschwerdt and Andrei Bykov
who helped us to revise the manuscript and corrected some misunderstanding of the model.
Many thanks to the anonymous reviewer, who sent us a detailed comment and suggestion which help us to improve the manuscript significantly.
Many thanks to the editor, Damiano Caprioli, and to the editorial team for handling the manuscript.
It was a pleasure for us to work in a benevolent and warm atmosphere.


\end{document}